\documentclass[12pt]{article}
\usepackage{epsfig}
\usepackage{amssymb}
\usepackage{amsmath}
\usepackage{array}
\newcommand{\m}{\mathbf}
\newcommand{\beq}{\begin{equation}}
\newcommand{\eeq}{\end{equation}}
\newcommand{\beqa}{\begin{eqnarray}}
\newcommand{\eeqa}{\end{eqnarray}}
\begin{document}

\title{THE RESONANCE THEORY OF PROTON AND ALPHA DECAY FROM HOT SOURCES}
\maketitle

\begin{center}
\author{F.F.Karpeshin$^{1,2,3}$, \and G. La Rana$^2$, \and
Emanuele Vardaci$^2$, \\
 \and Augusto Brondi$^2$,  \and Renata Moro$^2$, \and
S.N.Abramovich$^4$,  \and and V.I.Serov$^4$  }
\end{center}
{\small\it
\begin{itemize}
\item[$^1$]Fock Institute of Physics, St. Petersburg State University, RU-198504
St. Petersburg, Russia  \vspace*{-3mm}
\item[$^2$]Istituto Nazionale di Fisica Nucleare, and
    Dipartimento di Scienze Fisiche dell'Universita di Napoli, Italy    \vspace*{-3mm}
\item[$^3$]Departamento di Fisica, University of Coimbra, P-3004-516 Coimbra,
Portugal    \vspace*{-3mm}
\item[$^4$]Russian Federal Nuclear Centre VNIIEF, RU-607190 Sarov,
Nizhny Novgorod Region, Russia
\end{itemize}  }

\begin{abstract}
The consecutive microscopic solution is presented of the problem of
tunneling of a particle through a potential barrier. The method is
applied to the $\alpha$- and proton decay of compound systems
formed in fusion reaction.  Appearance of the peaks in the spectrum of
emitted particles is predicted. The peaks correspond to quasistationary
states inside the potential barrier.
\end{abstract}

\newpage

\section{Introduction}

Prefission emission of alphas and light charged particles provides with a
source of information about the time scale in the fusion-fission reaction.
On the other side, theoretical description of the processes  is usually made
in terms of the inverse cross-section. Such an approach assuming time
reversibility of the process leaves out of the scope a possibility of its
profound experimental check. This is in contrast with a number of indications
that violations of the reversibility may arise due to back-transparency
of the inner slope of the potential barrier in the ingoing channel [1],
or different response of the nuclear surface on the interaction with the
emitted and the same incoming particle, or due to temperature effects on
the barrier distribution [2], as in the ingoing channel experimental fit
of the optical model parameters is only possible for cold nuclei [3].

    Moreover, traditional decay theory deals with tunnelling
through a barrier of a particle which is in the quasistationary
state. This does not involve important cases when a virtual proton
or cluster is between the quasistationary states, as in alpha decay
from compound systems formed in fusion reaction.

    Our approach allows one to calculate the decay width at any
energy of the emitted particle. Strong resonance effects are,
specifically, predicted in alpha spectra from compound systems produced
in heavy-ion collisions.

\section{Formalism}

\subsection{$\alpha$ decay widths.}

Let $\Psi(\m r_1,\ldots, \m r_A)$ be a wave function of the source nucleus
with the mass number A. First, it can be expressed in terms of the
channel wave-function basis, as products of the wave-function of the
daughter nucleus $\varphi_n(\m r_1,\ldots,\m r_{A-4})$
and the $\alpha$-particle w.f. $\chi_k^{(L)}(\m r_{A-3},\ldots,\m r_A)$:
\beq
\begin{split}
\Psi\,=\,\sum_{L=0}^{L_0}\sum_{nk}C_{nkL}\varphi_n(\m r_1,\ldots,\m r_{A-4})
\chi_k^{(L)}(\m r_{A-3},\ldots,\m r_A)\equiv  \\
\equiv\,\sum_{L=0}^{L_0}\sum_{n}C_{nL}\varphi_n(\m r_1,\ldots,\m r_{A-4})
\eta_L^{(f)}(\m R;\m r_{A-3},\ldots,\m r_A)\;\;,
\end{split}
\eeq
where we selected the angular momentum $L$ of the relative motion of
the $\alpha$ particle in the nucleus. $\eta_L^{(f)}$ may be treated
as a wave-function of the relative motion of the $\alpha$ cluster in
the mother nucleus which evidently turns out to depend on the relative 
coordinate $\m R$. In simple cases
of pure configuration the expansion coefficients $C_{nkL}$ are reduced
to genealogical coefficients.

Let then the nucleus make a transition $i\rightarrow f$. As a result,
in the exit channel we observe  the system in a state which is described by a
wavefunction as a superposition
of the plane wave and ingoing spherical wave \cite{land} at large $\alpha$-nucleus
distances $R$:
\beq
\psi_{f\bf p}({\bf r}_1,\ldots,{\bf r}_A)
{\quad\raisebox{-9pt}{${\displaystyle\sim}\atop{\mbox{\rule{0pt}{11pt}}\scriptstyle R\; \to \;\infty}$}\quad}
\varphi_f(\m r_1,\ldots, \m r_{A-4})g_{\bf p}(\m r_{A-3},\ldots,\m r_A)\;\;,
\eeq
\beqa
g_p({\bf r}_{A-3},\ldots,{\bf r}_A)
{\quad\raisebox{-9pt}{${\displaystyle\sim}\atop{\mbox{\rule{0pt}{11pt}}\scriptstyle R\; \to \;\infty}$}\quad}
\left[e^{i{\bf pR}}+\frac{A(\vartheta,\varphi)}{R}e^{-ipR}\right]
\xi(\m r_{A-3},\ldots, \m r_A)\;\;;
\\
{\cal F}_{\bf p}({ \m R})\equiv
\left[e^{i{\bf pR}}+\frac{A(\vartheta,\varphi)}{R}e^{-ipR}\right]\;\;. \nonumber
\eeqa
In eq. (3), $g_{\bf p}$ is the channel wave-function, which is the
eigen function of the $\alpha$-nucleus Hamiltonian with an
appropriate mean-field single particle potential $U_\alpha(R)$:
\beq
(H-\varepsilon_{\bf p}){\cal F}_{\bf p}=0\,,
\eeq
\beq
\varepsilon_p=p^2/2M_\alpha\,.
\eeq
Furthermore, taking into account the asymptotics (3), the wavefunction
${\cal F}_{\bf p}({\m R})$ can be expressed in terms of the spherical
harmonics in a usual way:
\beq
{\cal F}_p({\bf R})=\sum_{\ell=0}^{\infty}i^\ell(2\ell+1)e^{i\delta_\ell}
R_{p\ell}(R)Y_{\ell m}(\theta,\varphi)\,.
\eeq

To find a transition amplitude, one has to change to the coordinate
system of the exit channel $|f{\bf p}\rangle $. The transformation  of
eq. (1) then conventionally reads as
\beq
\Psi=\sum_{\bf p}\langle \psi_{f{\bf p}}|\Psi\rangle \psi_{f{\bf p}}\,,
\eeq
the re-expansion coefficients giving the transition amplitude under
consideration. This way is similar to that found by Migdal when solving
his classical problem of shake of an atom in $\beta$ decay \cite{land}.
Substituting eqs. (1) together with (2), (3) and (6) into
eq. (7), we arrive at the following expression for the transition
amplitude:
\begin{multline}
M_{f{\bf p}}\,=\,\sum_{\ell=0}^{L_0}C_{f\ell}\,i^\ell e^{i\delta_\ell}
Y_{\ell m}(\theta,\varphi)\times
\\
\times\langle R_{p\ell}(R)Y_{\ell m}(\theta,\varphi)\xi({\bf r}_{A-3}-
{\bf R}_1,\ldots,{\bf r}_A-{\bf R})|\eta_\ell^{(f)}({\bf r}_{A-3},
\ldots,{\bf r}_A)\rangle \equiv \\
\equiv\,\sum_{\ell=0}^{L_0}C_{f\ell}\langle {\bf p}\xi|f\xi\rangle _\ell
\,i^\ell e^{i\delta_\ell}Y_{\ell m}(\theta,\varphi)\,\;.
\end{multline}
Taking into account that the wave functions $\cal F_{\bf p}$ are
normalized at 1 particle in a unit volume, with the flux $v\equiv
{\bf p}/M_\alpha$, we obtain from (8) the following expression for the
decay probability per a unit time:
\beq
\Gamma_{\bf p}\equiv\frac{d^3W}{d^3p}=|M_{f{\bf p}}|^2 \, v\,\;.
\eeq
Inserting $M_{f{\bf p}}$ from eq. (8) into eq. (9) and integrating over
all the angles of emission within $4\pi$, we arrive at the following
final expression for the decay width:
\beq
\Gamma_\alpha\equiv\frac{dN}{d\varepsilon_\alpha}=4\pi M_\alpha^2 \, v
\sum_{k}\sum_{\ell=0}^{L_0}|C_{f\ell}|^2 \, |\langle p|f\rangle _\ell|^2\,.
\eeq

\section{Method of numerical solution. Eigenvalues}

\begin{figure}[!b]
\centerline{
\epsfxsize=10cm
\epsfysize=8cm
\epsfbox{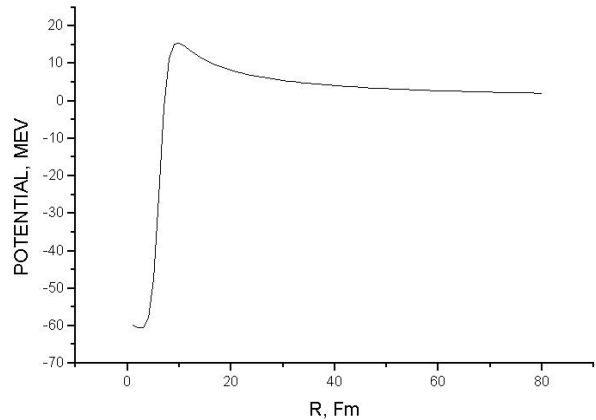} }
\caption{$\alpha$-nucleus potential for the system of $ ^{131}$La + $\alpha$
with the zero angular momentum.}
\label{WSPOT}
\end{figure}

$\alpha$-nucleus potential is characterized by a Coulomb barrier,
which is high enough, to form quasibound states inside the barrier
(Fig. \ref{WSPOT}).

These would be usual eigenstates, if the barrier were infinitely
broad. The values, however, go over the resonances on the continuum
background, whenever the penetrability of the barrier is taken into
account. Coupling to the continuum causes the energy shift and
broadening of the eigenstates. Affected eigenvalues can be determined
as follows.

The Schr\"{o}dinger equation for an $\alpha$-particle in the field of a
nucleus reads as follows:
\beq
\left\{-\frac{1}{2m}\left[\frac{d^2}{dr^2}-\frac{L(L+1)}{r^2}\right]
+V(r)\right\}\Psi=E\Psi\,,
\eeq
with the potential
\beq
V(r)=V_{SW}(r)+V_C(r)\,,
\eeq
\beq
V_{SW}=\frac{-V_0}{1+\exp\left(\dfrac{r-c}{a}\right)}\,,\label{A3}
\eeq
and the Coulomb potential was taken into account  as due to the
sharp-edge charge distribution:
\beqa
V_C(r) \, = \,   \left\{
    \begin{array}{l@{\hspace{2cm}}l}
\frac{\alpha Z}{2R_0}\left[3-\left(\frac{r}{R_{ 0}}
\right)^2\right]
\, &  \mbox{for } r<R_0\,,  \\  
\frac{\alpha Z}{r}\,
    &  \mbox{for }  r\geq R_0\,,
 \end{array}
   \right.
\eeqa
with $R_0$ being the nuclear radius.

On the radius segment between the origin $r=0$ and the first turning
point $R_{c1}$: $0<r\leq R_{c1}$, eq. (1) was integrated numerically
with the initial condition
\beq
\Psi(r)
{\quad\raisebox{-9pt}{${\displaystyle\sim}\atop{\mbox{\rule{0pt}{11pt}}\scriptstyle r\; \to \;0}$}\quad}
{r^L}\;\;.
\eeq

    General solution of the Schr\"{o}dinger equation under the barrier is a
linear combination of the two linearly independent solutions. One of
them exponentially vanishes, and the other exponentially increases with
increasing $R$. The coefficients can be obtained by sewing the
functions at the internal turning point. In principle, the eigenvalues
may be obtained from a condition that the coefficient for the
exponentially increasing solution vanishes. In the first approximation,
this can be achieved by sewing to the Airi function \cite{land}.
Actually, the eigen solutions were obtained by numerical integration
from the external turning point $R_{c2}$ towards the internal one, with
somewhat an arbitrary initial condition
\beq
y(R_{c2})=1\,,\qquad y'(R_{c2})=-0.25 \label{A6}\,.
\eeq
The derivative in eq. (6) is negative, as the solution is assumed to
exponentially decrease under the barrier.
In the course of integration, only the right solution survives, 
which exponentially increases with decreasing $R$  under the barrier,  
the other exponentially vanishes, in so far that the eigenvalue obtained
practically very weakly depends on the concrete numbers in eq. (\ref{A6}).

In general case, the fundamental set was obtained by numerical
integration from $R_{c2}$ to $R_{c1}$ with two different initial
conditions:
\beq
y(R_{c2})=1\,,\qquad y'(R_{c2})=\pm 1\,. \label{A7}
\eeq
After sewing at $r=R_{c1}$, the resulting solution increases under the
barrier (see below Figs. \ref{two} and \ref{three}) if not an eigenstate,
in contrast
with the behavior of each of the fundamental solutions. This
demonstrates mathematical correctness of the method. For the numerical
integration, the Runge-Kutta-Nystr\"{o}m method was used. The Shtermer
method was also tried, with essentially the same results. Behind the
barrier, the both solutions oscillate.

\section{Results and discussion}

\begin{figure}[!t]
\centerline{
\epsfxsize=10cm
\epsfysize=8cm
\epsfbox{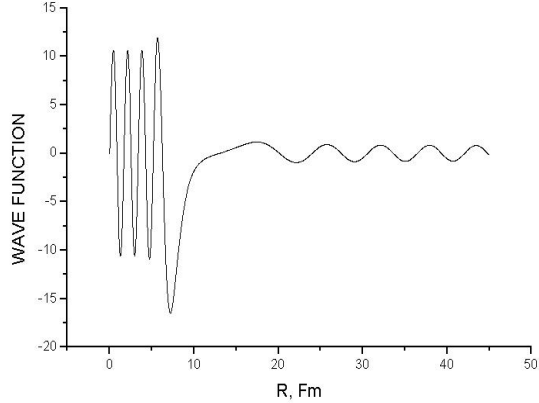}
}
\caption{$\alpha$ wavefunction for the system of $ ^{131}$La + $\alpha$ with the
$\alpha$ energy of 10.79 MeV.}
\label{one}
\end{figure}

\begin{figure}[!b]
\centerline{
\epsfxsize=10cm
\epsfysize=8cm
\epsfbox{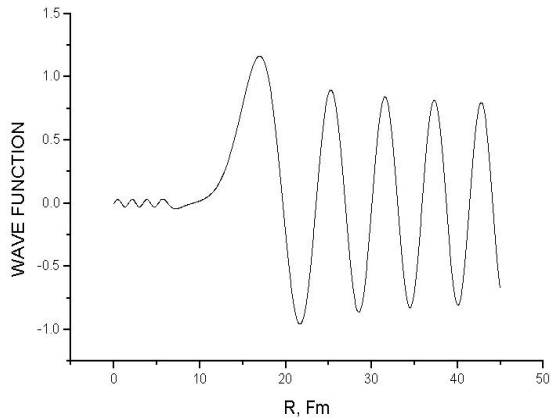} }
\caption{$\alpha$ wavefunction for the system of $ ^{131}$La + $\alpha$,
with the $\alpha$ energy of 11 MeV.}
\label{two}
\end{figure}

\begin{figure}[!t]
\centerline{
\epsfxsize=10cm
\epsfysize=8cm
\epsfbox{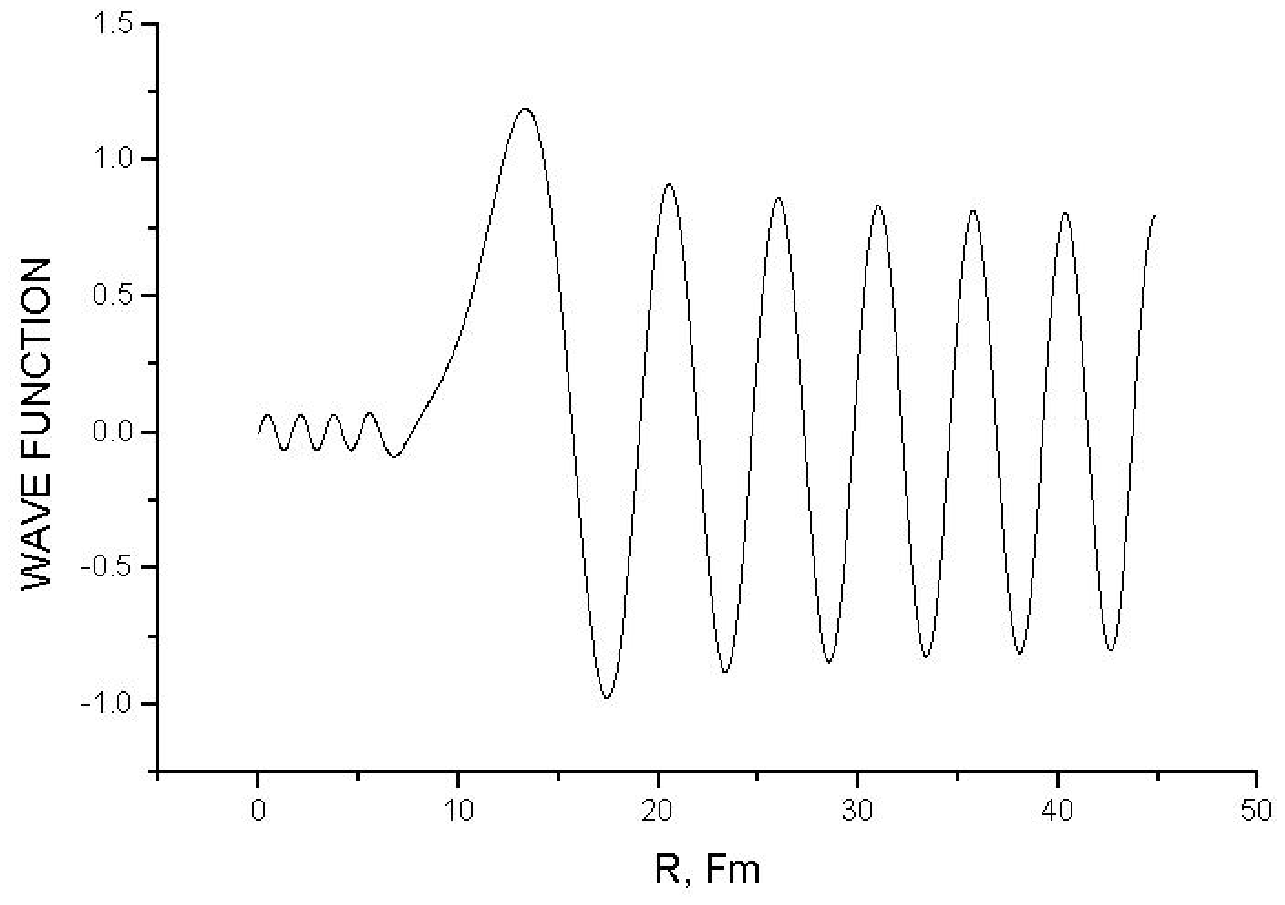} }
\caption{$\alpha$ wavefunction for the system of $ ^{131}$La + $\alpha$,
with the $\alpha$ energy of 14 MeV.}
\label{three}
\end{figure}

Calculations were performed with the Saxon-Woods potential (\ref{A3}),
with the parameters $V_0=100$ MeV, $s=2.3$ Fm, $c=1.2 A^{1/3}$ Fm.
Representative wavefunctions for various energies are presented in
Figs. \ref{one}--\ref{three} for the system $\alpha\,+\, ^{131}$La,  $\;\;L=0$.

Fig. \ref{one} answers the eigenvalue of $E_\alpha=10.79$ MeV. Corresponding
wavefunction has a large amplitude inside the barrier. Therefore, the
overlapping integral is also expected to be large in this case.

In Figs. \ref{two} and \ref{three}, we present the wavefunctions aside the resonance,
for the energies of 11 and 14 MeV, respectively. The wavefunctions are
normalised at $\delta( p-{p'})$. These figures are in drastic contrast
with the resonance one, presented in Fig. \ref{one}. The amplitude of the
wavefunction within the barrier is much smaller than outside. As a
result, the overlapping integral is expected to be small in the
nonresonance case, depressing nonresonance $\alpha$ decay.

\begin{figure}[!b]
\centerline{
\epsfxsize=10cm
\epsfysize=8cm
\epsfbox{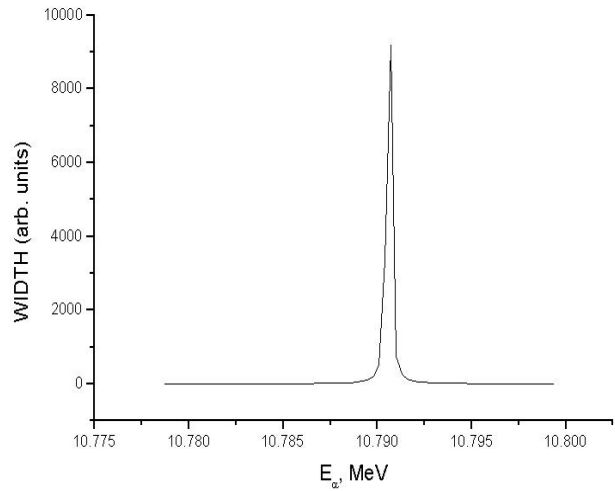} }
\caption{
Profile of the $\alpha$ decay line in an anticipated sub-barrier spectrum,
half-width being around 200 eV. (The system is $^{131}$La + $\alpha$, for the
alphas emitted with L = 0.) }
\label{four}
\end{figure}
    Finally, in Fig. \ref{four} we present  the calculated line
$\Gamma(E)$. That has a typical resonance shape with a half-width of around
200~eV. Therefore, spectrum of the subthreshold $\alpha$ particles turns out to
be modulated, directly indicating the resonance states inside the
barrier. Confirmation of this effect in experiment would really mean
discovering new physics.

In heavy-ion collisions, this effect may be smoothed by mixed
multipolarities. The effect must also manifest itself in usual $\alpha$- or
proton decay, specifically,  of nuclei far from the drip line.
In this case, set of the
allowed $L$ values is usually not large. Moreover, a partial wave with
a certain $L$ may make predominant contribution, which can be exploited
for direct check of the theory presented herein.
This study is planned to be made separately, in due course.

    One of us (FFK) would like to thank  H.~Valliser for many
fruitful discussions at the University of Coimbra and afterwards,
and Joao da Providencia for his support. Actually, this investigation was
started in 1999 within the framework
of the research undertaken in RFNC---VNIIEF, Sarov \cite{sarov}.
That was resumed during visiting the University of Naples, and continued
using a support from the Defense Threat Reduction Agency
(USA) under contract No.  DTRA01-01-P-0134,
and Russian Foundation for Basic Research under grant No. 02-02-17117.


\end{document}